# Thermodynamics and Relativity:
## Possible Consequences of their Close Link

Jean-Louis Tane (tanejl@aol.com). May 2008
Formerly with the Department of Geology, University Joseph Fourier, Grenoble, France

**Abstract**: The first part of this paper is a summary of a hypothesis previously advanced, suggesting the existence of a close link between thermodynamics and relativity. The second part is a preliminary comment about some possible consequences in the fields of physics, astronomy and biology.

## 1. Reminder of the reasons suggesting a link between thermodynamics and relativity

As proposed in previous papers [1, 2], let us consider a thermodynamic system defined as a given amount of gas contained in a vessel. We suppose that it can exchange work (dW) and heat (dQ) with the surroundings

If the gas evolves from an initial state $(P_1, V_1, T_1)$ to a final state $(P_2, V_2, T_2)$, the thermodynamic analysis of the process leads to three pieces of information.

### 1.1 First information

Concerning the work exchange, whether the process is irreversible or reversible, the equations we respectively have to use are:

$$dW_{irr} = -P_e dV \qquad (1)$$

$$dW_{rev} = -P_i dV \qquad (2)$$

where $P_e$ means "external pressure" and $P_i$ "internal pressure".

An immediate consequence is that, for a given value of dV, the difference between $dW_{irr}$ and $dW_{rev}$ takes the form:

$$dW_{irr} - dW_{rev} = dV(P_i - P_e) \qquad (3)$$

Observing that dV (which means $dV_{gas}$) is positive when $P_i > P_e$ and negative when $P_i < P_e$, it can be noted that the term $dV(P_i - P_e)$ is always positive, so that we have in all cases:

$$dW_{irr} > dW_{rev} \qquad (4)$$

Another writing of eq. 4 is therefore:

$$dW_{irr} = dW_{rev} + dW_{add} \qquad (5)$$

where $dW_{add}$ means $dW_{additional}$ and has a positive value.

## 1.2 Second information

In the classical conception of the first law of thermodynamics, it is admitted that when a system passes from a state A to a state B, its change in internal energy has the same value, whatever the level of irreversibility of the process. This proposition is expressed through the equality:

$$dU_{irr} = dU_{rev} \qquad (6)$$

For a process including an exchange of both thermal and mechanical energy (thermo-mechanical process) the detailed expression of dU is:

$$dU = dQ + dW \qquad (7)$$

Depending on whether the process is reversible or irreversible, eq. 7 takes the respective forms:

$$dU_{rev} = dQ_{rev} + dW_{rev} \qquad (8)$$

$$dU_{irr} = dQ_{irr} + dW_{irr} \qquad (9)$$

To conciliate the condition $dU_{irr} = dU_{rev}$ (eq. 6) with the condition $dW_{irr} > dW_{rev}$ (eq. 4), the only possible solution seems to be given by the proposition:

$$dQ_{irr} < dQ_{rev} \qquad (10)$$

## 1.3 Third information

According to the second law of thermodynamics, a change in entropy is linked to a change in heat by the equation:

$$dS = dQ/T + dS_i \qquad (11)$$

where $dS_i$ has a positive value for an irreversible process and a zero value for a reversible process.

Eq.11 has the dimension of an entropy, but takes the dimension of an energy if it is written under the form:

$$TdS = dQ + TdS_i \qquad (12)$$

In this last equation, the term $TdS_i$ is necessarily positive, since this condition is true for both $dS_i$ (as recalled above) and T (which is an absolute temperature).

As a conseqence, we get the inequality:

$$dQ < TdS \qquad (13)$$

Comparing eq. 10 and 13, we are tempted to consider as evident that $dQ_{irr}$ (in eq.10) corresponds to dQ (in eq.13) and $dQ_{rev}$ to TdS.

Although this conception is the one classically admitted, it is not totally convincing. Indeed, the precise meaning of eq. 11 is:

$$dS = dQ/T_e + dS_i \qquad (14)$$

so that the precise meaning of eq. 12 is itself:

$$T_e dS = dQ + T_e dS_i \qquad (15)$$

The term $T_e dS_i$ being positive (for the same reason as that already noted about $TdS_i$), the conclusion we are led to is:

$$T_e dS > dQ \qquad (16)$$

Remembering that $dQ_{irr} = T_e dS$ and $dQ_{rev} = T_i dS$ (in the same manner as $dW_{irr} = - P_e dV$ and $dW_{rev} = - P_i dV$), it seems that the significance of eq.16 is

$$dQ_{irr} > dQ_{rev} \qquad (17)$$

rather that the reverse.

Therefore, as already done with eq. 5 which refers to work, eq. 17 can also be written:

$$dQ_{irr} = dQ_{rev} + dQ_{add} \qquad (18)$$

where $dQ_{add}$ has a positive value.

Obviously, the result given by eq. 17 and 18 is not in accordance with eq. 10, so that we are confronted to another problem which requires a solution.

Observing that eq. 5 and 18 have the respective forms:

$$dW_{irr} = dW_{rev} + dW_{add} \qquad (5)$$

$$dQ_{irr} = dQ_{rev} + dQ_{add} \qquad (18)$$

where both $dW_{add}$ and $dQ_{add}$ are positive, it can be suggested that the interpretation usually admitted for first law of thermodynamics, that is the equality:

$$dU_{irr} = dU_{rev} \qquad (6)$$

needs to be substituted by the inequality:

$$dU_{irr} > dU_{rev} \qquad (19)$$

For a better parallel with eq. 5 and 18, eq.19 can also be written:

$$dU_{irr} = dU_{rev} + dU_{add} \qquad (20)$$

where $dU_{add}$ has a positive value.

Eq. 20 is an extended form of the first law of thermodynamics, which takes into account the problem evoked above.

The question raised by this equation is evidently the origin of the energy designated as $dU_{add}$. The explanation advanced in previous papers [1, 2] suggests a disintegration of mass occurring within the system, according to the Einstein mass-energy relation $E = mc^2$. In this conception, $dU_{add}$ is considered as having the significance:

$$dU_{add} = dE = -c^2 dm \qquad (21)$$

the sign minus corresponding to the fact that a decrease in mass results in an increase in energy and conversely.

Due to this reference to the mass-energy relation, another writing of eq. 20 is:

$$dU_{irr} = dU_{rev} - c^2 dm \qquad (22)$$

which appears as a general connection between thermodynamics and relativity, covering the the first and second laws.

**2. Possible consequences of the link between thermodynamics and relativity**

Keeping in mind that, at present, this link is nothing but an hypothesis, some preliminary consequences can be expected. The three ones, briefly evoked below, refer to physics, astronomy and biology.

**2.1. Concerning physics**

The conceptual difficulties often encountered in thermodynamics are somewhat reduced when the considerations just evoked are inserted in the equations dealing with the second law. This can be done as follows:

For a reversible process, the precise meaning of the classical equation $dS = dQ/T$ is:

$$dS_{rev} = dQ_{rev}/T_i \qquad (23)$$

where $T_i$ refers to the internal temperature of the considered system.

For an irreversible process, it has been already recalled that the meaning of the classical equation:

$$dS = dQ/T + dS_i \qquad (11)$$

is:

$$dS = dQ/T_e + dS_i \qquad (14)$$

Due to the analogy admitted above between eq. 16 and 17, the complementary information now available is that the significance of eq. 11 and 14 is more precisely:

$$dS_{irr} = dQ_{rev}/T_e + dS_i \qquad (24)$$

This expression is an entropy equation, its translation in an energy equation is:

$$T_e dS_{irr} = dQ_{rev} + T_e dS_i \qquad (25)$$

Knowing that entropy is a state function (as well as the volume), when a system passes from a state 1 to a state 2, the variation dS is the same, whatever the level of irreversibility of the process. As a consequence, we have $dS_{irr} = dS_{rev}$ so that combining eq. 23 and 24 leads to the general result:

$$dS_i = dQ_{rev}\,[1/T_i - 1/T_e] \qquad (26)$$

If $T_e > T_i$, the term in brackets is positive and the system receives heat from the surroundings, so that $dQ_{rev}$ is positive. Therefore $dS_i$ is positive

If $T_e < T_i$, the term in brackets is negative and the system provides heat to the surroundings, so that $dQ_{rev}$ is negative. Therefore $dS_i$ is positive.

Although this result is a well-known data of the thermodynamic theory, the important point of the reasoning just adopted is that in equation 14, the term dQ represents $dQ_{rev}$, not $dQ_{irr}$. This peculiarity explains that it can be factorized in eq 26, which is a synthesis of eq. 23 (reversibility) and 24 (irreversibility), taking into account the equality $dS_{irr} = dS_{rev}$.

An interesting extension concerns the case of a system defined as isolated, composed of two parts (designated 1 and 2) which exchange heat, because their initial temperatures are different ($T_1$ and $T_2$).

Applying eq. 26 to part 1 and part 2 gives respectively:

$$dS_{i1} = dQ_{rev1}\,[1/T_1 - 1/T_2] \qquad (27)$$

$$dS_{i2} = dQ_{rev2}\,[1/T_2 - 1/T_1] \qquad (28)$$

Since $dQ_{rev2} = -\,dQ_{rev1}$, eq. 28 can also be written $dS_{i2} = dQ_{rev1}\,[1/T_1 - 1/T_2]$, showing that we have the equality:

$$dS_{i1} = dS_{i2} \qquad (= dS_i) \qquad (29)$$

where $dS_i$ is a general designation (used in eq. 35 below) for this common value.

As a consequence, if the formula applied to part 1 and part 2 is eq. 25, we get:

$$T_2 dS_{irr1} = dQ_{rev1} + T_2 dS_{i1} \tag{30}$$

$$T_1 dS_{irr2} = dQ_{rev2} + T_1 dS_{i2} \tag{31}$$

whose meanings are respectively:

$$dQ_{irr1} = dQ_{rev1} + dQ_{add1} \tag{32}$$

$$dQ_{irr2} = dQ_{rev2} + dQ_{add2} \tag{33}$$

Remembering that $dQ_{rev\,2} = - dQ_{rev\,1}$ and $dS_{i2} = dS_{i1}$, the result obtained for the whole system is therefore:

$$dQ_{irr.syst} = dQ_{rev.syst} + dQ_{add.syst} \tag{34}$$

that is: 
$$dQ_{irr.syst} = 0 + dS_i (T_1 + T_2) \tag{35}$$

where $dS_i (T_1 + T_2)$ has always a positive value.

Similarly, if we consider an isolated system composed of two gaseous parts which exchange work (because their initial pressures $P_1$ and $P_2$ are different), it can be derived from eq. 5 that:

$$dW_{irr1} = dW_{rev1} + dW_{add1} \tag{36}$$

$$dW_{irr2} = dW_{rev2} + dW_{add2} \tag{37}$$

The result for the whole system is therefore:

$$dW_{irr.syst} = dW_{rev.syst} + dW_{add.syst} \tag{38}$$

that is 
$$dW_{irr.syst} = 0 + dV_1 (P_1 - P_2) \tag{39}$$

Beyond the fact that the system is concerned with an internal disappearance of the pressure gradient (or the temperature gradient in the case of an exchange of heat), is it possible to imagine a symptom, visible by an external observer, indicating that a process occurs within the system? The next paragraph, which deals with astronomy, is devoted to a very preliminary comment about this matter.

### 2.1. Concerning astronomy

Among the fundamental tools introduced by Newton and closely connected to his genaral law of gravitation:

$$F = GMm/R^2 \qquad (40)$$

is the equation:

$$F = m\gamma \qquad (41)$$

often called the second law of motion.

Referring to the gravitational field of the Earth, eq. 41 takes the particular form:

$$F = mg \qquad (42)$$

From eq. 42 has been derived the formula:

$$E = mgh \qquad (43)$$

The significations of eq. 40, 41, 42 and 43 are well known.

Admitting as a first convention that $h = 0$ at the Earth's surface, and therefore that $E = 0$, a basic use of eq. 43 consists in calculating the change in potential energy of an object of mass $m_1$, whose altitude varies from $h_1$ to $h_2$.

The answer usually given is:

$$\Delta E = mg\,(h_2 - h_1) \qquad (44)$$

which contains the implicit idea that m is constant and that g is constant too (at least when the difference of altitude remains small).

Now let us consider that the object is a satellite orbiting around the Earth and concerned with internal exchanges in energy. According to the hypothesis previously advanced, these internal processes lead to a decrease of its mass, so that this object passes from a state 1 (mass $m_1$) to a state 2 (mass $m_2$), obeying the condition $m_2 < m_1$.

How can we reconcile this peculiarity with the law of conservation of energy?

A possible answer is to write that the gravitational energies of the object, in the states 1 and 2, are respectively:

$$E_1 = m_1 g_1 h_1 \qquad (45)$$

$$E_2 = m_2 g_2 h_2 \qquad (46)$$

Having $E_2 = E_1$ (law of conservation of energy), we get:

$$h_2 = (m_1 g_1 h_1) / m_2 g_2 \qquad (47)$$

Admitting, as a first hypothesis, that $g_1$ and $g_2$ can be considered equal, the value $h_2$ is simply:

$$h_2 = (m_1 h_1) / m_2 \tag{48}$$

which, combined with the condition $m_2 < m_1$, leads to the conclusion:

$$h_2 > h_1 \tag{49}$$

The significance of eq. 49 is that the distance h separating the satellite from the Earth tends to increase, due to the exchanges of energies occurring within the satellite (those occurring within the Earth have, of course, the same kind of effect).

This situation is strengthened if we take into account that "g" goes decreasing with the distance, because having $g_2 < g_1$, the value obtained for $h_2$ with eq. 47 is greater than the one obtained with the simplified eq. 48.

In other words, eq. 49 means that the system defined as the set "Earth + satellite" is in expansion and eq. 47 brings the precision that this expansion is accelerated.

Due to the fact that the validity of eq. 41 is general, it seems that the same reasoning can be extended to the relation between a planet and a star, or between two stars or two galaxies. In any case, the suggested idea is that the exchanges of energy which occur inside an object result in a decrease of its mass and therefore in an increase of the distance separating this object from those with which it is linked by a gravitational field. From this point of view, if the connection suggested between eq. 22 and eq. 47 is recognized as valid, it may be the sign of a rather simple convergence between the theory of Newton (general law of gravitation) and the theory of Einstein (mass-energy relation). Note that the idea of an increasing distance separating systems linked by gravitation seems to be in accordance with Hubble's law [3] and to give a possible answer to the so-called "reversibility paradox" [4].

A correlative point which can be noted (but just qualitatively) is that the greater the level of irreversibility of an exchange in energy, the greater the corresponding decrease in mass, (and therefore the increase in distance just evoked), but the shorter the time necessary for the process to be achieved. The idea which seems emerging from these considerations is the possibility that the concepts of mass, time and distance are linked, although they appear to us, at first glance, as independent from one another.

### 2.1. Concerning biology

It is well known that there is a fundamental difference of behavior between living matter and inert matter and that the accordance of living bodies with the laws of thermodynamics seems better after their death than before.

Since these laws have been stated from experiments performed on systems made of inert matter (such as motors), it is not evident that they are directly applicable to systems made of living matter.

When thermodynamics is examined from the microscopic point of view, the internal increase in entropy of the macroscopic approach (i.e. the positive value of the term $dS_i$ in eq.11) is interpreted as an increase in disorder. With this convention, a system composed of two parts whose initial temperatures are $T_1$ and $T_2$, is considered as increasing in disorder as the difference of the temperatures goes decreasing, due to a natural exchange of heat between the two parts. At first glance, this convention looks arbitrary, but its logic is more easily understandable when the system considered is an artificial building, for example a castle. Its natural evolution being a progressive weathering and disintegration, it appears clearly as an increase in disorder.

As briefly mentioned above, it is mainly after its death that the behavior of a living organism presents a similar evolution. During its life, and particularly during the period of its germination and growth, it seems that it evolves towards an increasing order. The resulting question, which has been asked for a long time [5], is therefore: is it possible that the laws of thermodynamics are inverted when the system we refer to is made - at least partially - of living matter?

We have noted above that eq. 22, whose expression is:

$$dU_{irr} = dU_{rev} - c^2 dm \qquad (22)$$

covers the first and second laws and appears as a connection between thermodynamics and relativity.

In conventional thermodynamics, the mass of an isolated system is considered as invariable, even when this system is concerned with internal processes. This conception has been extended to closed systems, because the energy exchanged with the surroundings is not great enough to make measurable the correlative mass variation of the system. The best known example of such an extension is the principle of conservation of mass in chemical reactions, stated by Lavoisier.

In the conception summarized through eq. 22, the positive value of $dS_i$ is interpreted as the symptom of a positive energy (noted $T_e dS_i$ in eq. 15 and 25) whose corresponding change in mass is represented by the negative value of dm.

If the laws of thermodynamics are inverted for living organisms, the change that can be imagined is a positive value of dm instead of a negative value. From the practical point of view, this would mean that a living system transforms energy into mass, contrary to an inert system that transforms mass into energy.

According to a series of experiments performed a few years ago [6], it seems that positive changes in mass have been observed, concerning closed systems made of a mixture of living and inert matter. Referring to the importance of such data, their checking by further investigations would certainly be of great interest.

## 3. Conclusions

As already mentioned, several of the points discussed above have been introduced in previous papers [1, 2]. The main novelty is the hypothesis presented in section 2.1, suggesting a link between the general law of gravitation and the mass-energy relation. The insertion of the latter in the thermodynamic theory (eq. 22) is the key providing access to such openings.

**References**

[1]    J-L Tane, *Thermodynamics and Relativity: A Condensed Explanation of their Close Link*. arxiv.org/pdf/physics/0503106, March 2005.

[2]    V. Krasnoholovets and J-L Tane, *An extended interpretation of the thermodynamic theory, including an additional energy associated with a decrease in mass* Int. Journal Simulation and Process Modelling, Vol. 2, Nos. 1/2, 2006 67.

[3]    M. Longair, *Theoretical Concepts in Physics*. Cambridge University Press, Second edition, 2003. (pp. 502-504)

[4]    D. Cassidy, G. Holton and J. Rutherford, *Understanding Physics*, Springer-Verlag, New York, 2002 (pp. 320-322)

[5]    E. Schrödinger, "*What is Life* ?", New York 1944, Cambridge University Press (reprinted in 1992, with a foreword by Roger Penrose).

[6]    A. Sorli, *The Additional Mass of Life*. Journal of Theoretics, April/May 2002, Vol.4 No 2.